# Giant anomalous Hall conductivity in frustrated magnet $EuCo_2Al_9$

Sheng Xu[†,1,2,*] Jian-Feng Zhang[†,3] Shu-Xiang Li,[2] Junfa Lin,[4] Xiaobai Ma,[5] Wenyun Yang,[6] Jun-Jian Mi,[2] Zheng Li,[2] Tian-Hao Li,[2] Yue-Yang Wu,[2] Jiang Ma,[2] Qian Tao,[2] Wen-He Jiao,[1] Xiaofeng Xu,[1] Zengwei Zhu,[7] Yuanfeng Xu,[8] Hanjie Guo,[9,†] Tian-Long Xia,[4,10,‡] and Zhu-An Xu[2,11,12,§]

[1] School of Physics, Zhejiang University of Technology, Hangzhou 310023, China

[2] State Key Laboratory of Silicon and Advanced Semiconductor Materials & School of Physics, Zhejiang University, Hangzhou 310058, China

[3] Center for High Pressure Science & Technology Advanced Research, 100094 Beijing, China

[4] School of Physics and Beijing Key Laboratory of Opto-electronic Functional Materials & Micro-nano Devices, Renmin University of China, Beijing 100872, China

[5] Neutron Scattering Laboratory, Department of Nuclear Physics, China Institute of Atomic Energy, Beijing 102413, China

[6] State Key Laboratory for Mesoscopic Physics, School of Physics, Peking University, Beijing 100871, China

[7] Wuhan National High Magnetic Field Center and School of Physics, Huazhong University of Science and Technology, Wuhan 430074, China

[8] Center for Correlated Matter, School of Physics, Zhejiang University, Hangzhou 310058, China

[9] Songshan Lake Materials Laboratory, Dongguan, Guangdong 523808, China

[10] Key Laboratory of Quantum State Construction and Manipulation (Ministry of Education) & Laboratory for Neutron Scattering, Renmin University of China, Beijing 100872, China

[11] Hefei National Laboratory, Hefei 230088, China

[12] Collaborative Innovation Centre of Advanced Microstructures, Nanjing University, Nanjing 210093, China

The interaction between conduction electrons and localized magnetic moments profoundly influences the electrical and magnetic properties of materials, giving rise to a variety of fascinating physical phenomena and quantum effects. Here, we discover a giant anomalous Hall effect (AHE) in a frustrated Eu-based magnet, exhibiting a giant anomalous Hall conductivity (AHC) of $3.1 \times 10^4\ \Omega^{-1}\mathrm{cm}^{-1}$ and a remarkable anomalous Hall angle (AHA, $\tan\theta_H$) of 12 %—surpassing conventional mechanisms (either intrinsic or extrinsic) by two orders of magnitude. Combining magnetotransport, quantum oscillations, neutron diffraction and ab initio calculations, we establish that the giant AHC originates from fluctuating spin chirality skew scattering, generated by indirect Ruderman-Kittel-Kasuya-Yosida (RKKY) interactions of Eu-4*f* moments. Simultaneously, Hund's coupling of itinerant electrons and localized Eu-4*f* spins triggers giant exchange splitting, evidenced by temperature-dependent Fermi surface reconstruction. This work establishes a frustrated magnetic platform for engineering the AHE and elucidates the governing role of exchange interactions and spin textures in quantum transport, while also providing a framework for designing unconventional spintronic systems that harness emergent spin-texture dynamics.

[*]shengxu_phy@zju.edu.cn; [†]hjguo@sslab.org.cn; [‡]tlxia@ruc.edu.cn; [§]zhuan@zju.edu.cn

## Introduction

The anomalous Hall effect (AHE), a cornerstone of quantum transport in magnetic systems, has emerged as a transformative platform for spintronic technologies—from nonvolatile memory architectures to ultrasensitive magnetic sensors—by exploiting spin-polarized currents and emergent topological states. Noncoplanar spin textures with finite scalar spin chirality $(\langle S_i \cdot (S_j \times S_k) \rangle)$ generate giant Berry curvature through topologically protected band crossings [1–9], and collinear ferromagnets with spin-orbit-coupled band reconstructions exhibit analogous enhancements [10–12]. While intrinsic AHE driven by momentum-space Berry curvature offers robust quantization, its conductivity remains fundamentally constrained to values below $e^2/ha$ ($10^3$ $\Omega^{-1}$ $cm^{-1}$), limiting practical device efficiency. When the mean free path becomes shorter than the characteristic size of the spin texture, spatially slowly varying moments, such as those in skyrmions, induce a Berry phase in real space, giving rise to the topological Hall effect (THE) [13–16]. Extrinsic mechanisms such as skew scattering, though capable of exceeding this threshold, suffer from suppressed anomalous Hall angles (AHA, $\tan\theta_H=\sigma_{xy}/\sigma_{xx}$ ~ 0.1-1 %) due to ultrahigh longitudinal conductivity $\sigma_{xx} > 5\times10^5$ $\Omega^{-1}$ $cm^{-1}$ in conventional systems [17–19].

Recent breakthroughs reveal that geometrically frustrated spin systems provide an unprecedented design space to transcend these limitations [20–28]. A key discovery involves a paradigm-shifting mechanism: chiral spin clusters, even in disordered or thermally fluctuating regimes, enable a cooperative spin chirality skew scattering process that amplifies anomalous Hall conductivity (AHC) by orders of magnitude while maintaining large AHA—as exemplified in systems such as $KV_3Sb_5$ [22], MnGe [23], EuAs [24], $SrCo_6O_{11}$ [25], $PdCoO_2$ [26, 27] and $GdCu_2$ [28]. This phenomenon persists above magnetic ordering temperatures, a critical feature for room-temperature applications. However, the experimental realization of this mechanism remains underexplored; material systems that exhibit both giant AHC/AHA and unambiguous evidence of spin chirality skew scattering are still highly limited. The search for and confirmation of such behavior in a broader range of structurally well-defined and magnetically tunable frustrated lattices is therefore crucial. It will not only help establish a universal microscopic picture of the mechanism but also provide a solid materials foundation and physical principles for AHE-based spintronic applications—such as highly sensitive magnetic memory, low-power Hall sensors, and non-volatile logic devices.

In $EuCo_2Al_9$, which crystallizes in the centrosymmetric space group *P6/mmm*, the $Eu^{2+}$ ions form a frustrated triangular lattice. In this structure, the Dzyaloshinskii-Moriya interaction is forbidden by the inversion symmetry [29, 30]. Furthermore, the super-exchange interaction should also be ineffective in this system due to the lack of localized, polarizable anion states [31, 32]. Therefore, the interactions between Eu moments should be dominated by the indirect Ruderman-Kittel-Kasuya-Yosida (RKKY) interaction, mediated by itinerant electrons. This RKKY interaction, when combined with an applied magnetic field, can stabilize non-coplanar spin textures, thereby creating an ideal platform for exploring chiral spin-cluster-driven quantum transport. We report a giant AHE in this system, manifested by a giant AHC of $3.1\times10^4$ $\Omega^{-1}$ $cm^{-1}$ and a remarkable AHA of 12 %. These values, exceeding conventional mechanisms (either intrinsic or extrinsic) by two orders of magnitude, point to an origin in spin chirality skew scattering. This work establishes $EuCo_2Al_9$ as a frustrated magnetic platform for tailoring the AHE, while providing a framework for designing unconventional spintronic systems that exploit emergent spin-texture dynamics.

## Result

## Crystal structure, magnetic and electric properties

$EuCo_2Al_9$ single crystals, synthesized via a self-flux method (Methods), crystallize in the hexagonal space group *P6/mmm* (Fig. 1a). The structure comprises four distinct sublattices: triangular Eu layers, Co honeycomb networks, Al-1 kagome and Al-2 planes (Fig. 1**a**,**b**), creating a hierarchical geometry that intertwines frustration and symmetry. In the triangular Eu sublattice, a hallmark of geometric frustration, the magnetic structure is stabilized by RKKY interactions, whereas the non-magnetic Co and Al sublattices enable selective tuning of conduction electron pathways.

Figure 1**c** presents the temperature-dependent magnetic susceptibility ($\chi$) and inverse susceptibility ($1/\chi$), measured over the temperature range of 2–300 K. A Curie-Weiss analysis of the paramagnetic regime (100–300 K) yields an effective moment of 7.69$\mu_B$ and a Weiss temperature $\Theta_{CW}$ of 3.7 K, consistent with the phase transition observed at $T^*$~3.5 K. This magnetic behavior is absent in the isostructural nonmagnetic analogues $SrCo_2Al_9$ and $BaCo_2Al_9$ [33], providing direct evidence for its origin in $Eu^{2+}$ driven magnetism. This phase transition ($T^*$) is further highlighted in transport data: a sharp resistivity drop (see SI for more details) coincides with a λ-like specific heat anomaly (Fig. 1**e**), confirming a bulk magnetic transition. At 2 K, the $\boldsymbol{B} \parallel c$ isotherm magnetization exhibits two distinct plateaus (Fig. 1**c**, inset). The first plateau below around 2 T, with a magnetization of about 2.4 $\mu_B$/Eu (~1/3 of the full moment) that increases with field, signals a canted up-down-down (u-d-d) spin structure. Strikingly, even at 70 K, a magnetization of 2.86 $\mu_B$/Eu persists at 14 T, evidencing robust short-range spin correlations extending far above $T^*$.

Notably, in addition to the transition at $T^*$~3.5 K, the temperature dependence of specific heat reveals another distinct anomaly at a lower temperature $T^\dagger$~1.1 K, indicating a second phase transition. The magnetic contribution to the heat capacity of $EuCo_2Al_9$ at low temperatures, separated by subtracting the phonon-dominated signal of the non-magnetic analogue $BaCo_2Al_9$, is shown in the inset of Fig. 1**e**. The total magnetic entropy change tends to level off above 4 K, approaching a value of $R$ln8 (where $R$ is the ideal gas constant), consistent with an $S$=7/2 state. More strikingly, the entropy change from the base temperature to $T^\dagger$ is about 1/3 of the total value. This fractional liberation implies a hierarchical ordering process: two spins on the Eu triangular lattice develop antiferromagnetic order at $T^*$, while the third spin persists in a fluctuating paramagnetic state before undergoing ordering at $T^\dagger$. This proposed scenario will be directly confirmed in the forthcoming neutron diffraction studies.

## Neutron diffraction and magnetic structure.

In order to study the magnetic structure at low temperatures, neutron scattering experiments were performed. The nuclear structure measured at 10 K (see the SI for more details) can be described by the space group *P6/mmm*. At 2 K, additional magnetic reflections appear (Fig. 1**f**), which can be indexed with a propagation vector $k$=(1/3,1/3,0). To solve the magnetic structure, we performed irreducible representation (IR) analysis by using the program *BASIREPS* in the *FULLPROF* package [34]. In the space group *P6/mmm*, the magnetic reducible representations $\Gamma_{mag}$ for the 1$a$ (Eu) and 2$c$ (Co) sites are decomposed as:

$$\Gamma_{mag}(\mathrm{Eu}) = \Gamma_3 \oplus \Gamma_5 \quad (1)$$

$$\Gamma_{mag}(\mathrm{Co}) = \Gamma_2 \oplus \Gamma_4 \oplus \Gamma_5 \oplus \Gamma_6. \quad (2)$$

The basis vectors for these IRs can be found in Tab. S1. Since only one transition is observed around 3.5 K within the transport tested temperature range of 2-300 K, the Eu and Co sublattices should order under the same IR if there exists a coupling among them. However, a refinement based on the IR $\Gamma_5$ for both sublattices does not yield a satisfactory result. Alternatively, a good agreement can be achieved based on the IR $\Gamma_3$ for the Eu ions. As shown in Tab. S1, the spins are aligned along the *c*-axis, consistent with the magnetic susceptibility results. The moments at the lattice can be expressed as $\mathbf{m}_l = \mathbf{m}_0\cos(2\pi \boldsymbol{k}\cdot R_l + \varphi_0)$, where $R_l$ is the lattice vector and $\varphi_0$ is an initial phase. Usually, the initial phase $\varphi_0$ is not important in determining the magnetic structure. However, for some commensurate structures, as the case shown here, a choice of different $\varphi_0$ can result in completely different magnetic structures. If we assume $\varphi_0 = 0$, an u-d-d structure is obtained. However, in this case, the moment size of the down spin is half that of the up spin, which amounts to $7.6\mu_B$. This value is larger than the expected maximum moment size for $Eu^{2+}$ ($7\mu_B$). On the other hand, if we assume $\varphi_0 = \pi/2$, a 0-u-d structure is obtained with an equal size for the up and down spins. In this case, the best fit yields a moment size of $6.59\mu_B$. Although neutron powder diffraction measurements cannot distinguish between these two results, the argument based on the saturation moment supports the latter scenario. This is also consistent with the geometrical frustration and the observed 1/3 plateau phase. As is seen in the magnetic structure in Fig. 1**d**, the molecular field at the center of the hexagon is canceled out by the surrounding six moments, thus, the moment at the center keeps fluctuating in the paramagnetic state. Once a small magnetic field is applied, the moment at the center minimizes the energy by aligning with the magnetic field, leading to the 1/3 plateau phase. More generally, the moments can be canted if the Ising anisotropy is not strong, as is the case for $Eu^{2+}$, leading to a diminished plateau.

## Giant anomalous Hall effect.

Figure 2**a** presents the field-dependent magnetoresistance (MR) at different temperatures. The MR increases with the magnetic field and tends to saturate at high fields, a behavior that diminishes at temperatures above 70–85 K. At 2 K, a convex MR feature emerges around 2 T, suggesting a field-induced metamagnetic transition. The observed negative MR near zero field and 2 T is consistent with reduced magnetic scattering from moment reorientation and magnetic ordering. Figure 2**b** presents the field-dependent magnetization at various temperatures. In addition to the metamagnetic transition observed at 2 K, significant magnetization persists well above $T^*$. A value of $2.86\mu_B$/Eu is maintained at 70 K and 14 T, providing direct evidence for robust short-range spin correlations that extend far above the magnetic ordering temperature. Figure 2**c** exhibits the field-dependent Hall resistivity $\rho_{yx}$. As shown, the $\rho_{yx}$ develops a broad, hump-like anomaly below approximately 100 K and for magnetic fields below 7 T. The Hall resistivity $\rho_{yx}$ in magnetic systems is typically expressed as the sum of two contributions: $\rho_{yx} = \rho^N_{yx} + \rho^A_{yx}$, where $\rho^N_{yx}$ and $\rho^A_{yx}$ represent the normal and anomalous Hall resistivity. Isolating the anomalous contribution, $\rho^A_{yx}$, through mathematical decomposition of the total Hall signal is a procedure commonly employed in the analysis of magnetic systems [9, 35]. Through such analysis, we estimate $\rho^A_{yx}$ (see SI for more information), and obtain the $\sigma^A_{xy}$. At 2 K, $\sigma^A_{xy}$ reaches $\sim 3.1 \times 10^4\ \Omega^{-1}\ \mathrm{cm}^{-1}$, and gradually decreases to $0.7 \times 10^4\ \Omega^{-1}\ \mathrm{cm}^{-1}$ at 70 K (Fig. 2**e**). At low temperature, the ordinary Hall contribution is relatively small and exhibits little variation with temperature, enabling a robust separation and thus a reliable determination of the AHC.

To reliably separate the normal contribution, we adopt only the experimental data below 70 K and perform a scaling analysis of the $\sigma^A_{xy}$–$\sigma_{xx}$ relation from 2 K to 50 K, which classifies AHE mechanisms into three regimes: (i) the "dirty" regime ($\sigma_{xx}$ < $3\times10^3$ $\Omega^{-1}$cm$^{-1}$), (ii) the intrinsic Berry-curvature-dominated regime, and (iii) the "clean" regime ($\sigma_{xx}$ > $5\times10^5$ $\Omega^{-1}$cm$^{-1}$) [36]. $EuCo_2Al_9$ resides in the intrinsic regime ($\sigma_{xx} \sim 2\times10^5$ $\Omega^{-1}$cm$^{-1}$), yet its $\sigma^A_{xy}$ far exceeds the conventional quantum limit of $e^2/ha$ (dashed line in Fig. 2**e**; $a \approx 3.91$ Å). This discrepancy, combined with the observed large AHA ($\tan\theta_H \approx 12\%$ at 2 K), also rules out both intrinsic Berry curvature ($\sigma^A_{xy} \sim 10^2$ $\Omega^{-1}$cm$^{-1}$) and traditional skew scattering ($\tan\theta_H \sim 0.1-1\%$) as primary mechanisms. Instead, our findings align with the recently proposed model of skew scattering mediated by chiral spin clusters [20, 21], which amplifies both $\sigma^A_{xy}$ and AHA. This mechanism persists above magnetic ordering temperatures, suggesting a robust role of frustrated spin correlations. Figure 2**f** shows the scaling relation $\sigma^A_{xy} \propto \sigma_{xx}^n$ for $EuCo_2Al_9$ and other spin chirality skew scattering materials. In $EuCo_2Al_9$, $n$ is obtained to be 2.1 (see SI for more information). This places it close to systems like $KV_3Sb_5$ ($n \approx 2$) [22], $PdCoO_2$ ($n \approx 2$) and $GdCu_2$ ($n \approx 1.78$) [28], while differing markedly from MnGe ($n \approx 1$) [23]. This non-universality of $n$ may reflect differences in the resonance condition for spin chirality skew scattering [23]. Figure 3 displays the contour plot of $\sigma^A_{xy}$ across the $\boldsymbol{B}-T$ plane.

## First-principles calculations.

To investigate whether the giant AHE could originate from field-induced ferromagnetic (FM) exchange splitting—a mechanism that can reconstruct bands and generate large Berry curvature [37]—we calculated the band structures of $EuCo_2Al_9$ with and without Eu 4*f* moments. Although the ground state of $EuCo_2Al_9$ is antiferromagnetic (AFM), a field-induced FM-like alignment of Eu moments emerges at low temperatures and sufficiently high magnetic fields. Therefore, the FM calculation models this high-field limit of maximal exchange splitting, to assess whether the resultant band topology could support a large AHC.

The corresponding band structures and Fermi surfaces (FSs) are presented in Figs. 4**a**-**d**, respectively. In the absence of the Eu atomic local moment, the electronic states near the Fermi level predominantly originate from the Co atomic orbitals, with minor contributions from Eu and Al-1(Al-2) atoms, as detailed in our partial density of state (pDOS) analysis (Fig. 4**f**). In Fig. 4**b**, we carried out a regular FM order calculation for comparing with the non-magnetic case in panel Fig. 4**a**. We can find the bands with largest splitting mainly come from Eu-5*d* orbital, whose weights were also shown in Fig. 4**a**. Around the Fermi level, we can identify two Eu-5*d* orbital-related Fermi pockets: a large hole-type pocket near the Γ point and an electron-type pocket near the K point. Both of them exhibit substantial exchange splitting effects from the Eu-4*f* electrons, as demonstrated in Figs. 4**a** and 4**b**.

To understand the anomalous transport properties of this system, we calculated the AHC by applying the Kubo formula within the linear response framework, via the momentum space Berry curvature effects. The spin configuration of the Eu-4*f* electrons was aligned with *c*-axis collinear ferromagnetism, same with the calculation in Fig. 4**b**. Our results indicate that a pure out-of-plane easy axis can only yield a modest AHC of 10 $\Omega^{-1}$cm$^{-1}$ (Fig. 4**e**). This value is three orders of magnitude lower than our experimental observations, providing strong evidence that simple FM Zeeman-type exchange splitting alone cannot account for the intrinsically large AHC.

## Quantum oscillations.

Quantum oscillations provide critical insights into electronic band structure, where oscillation frequencies correlate with extremal FS cross-sections perpendicular to the applied magnetic field via the Onsager relation $F = (\hbar/2\pi e)A_F$ ($A_F$: FS cross-sectional area) [38]. Figure 5a displays the Shubnikov–de Haas (SdH) oscillations under B applied along c-axis. Fast Fourier transform (FFT) analysis of the oscillatory component $\Delta\rho_{xx}$ identifies four distinct frequencies: $F_\alpha$ = 219.6 T, $F_\beta$ = 349.7 T, $F_\gamma$ = 492.4 T and $F_\eta$ = 586.51 T. Strikingly, $F_\alpha$ and $F_\gamma$ increase with temperature, whereas $F_\beta$ and $F_\eta$ decrease (Fig. 5c), reflecting contrasting FS dynamics. These properties are consistent with de Haas–van Alphen (dHvA) oscillations in magnetization (see the SI for more details). From the Onsager relation, we derive Fermi wavevectors $k_F =(A_F /\pi)^{1/2}$, plotted in Figs. 5d,e. Effective masses, extracted via Lifshitz–Kosevich (LK) fits to temperature-dependent FFT amplitudes, range from 0.06$m_0$ ($F_\alpha$ ) to 0.08$m_0$ ($F_\gamma$ and $F_\eta$ ) (see the SI Tab. S2 for more details). These light effective masses rule out the Kondo-type band renormalization, in which the increasing delocalization of localized *f* electrons due to hybridization with itinerant *d* electrons upon cooling leads to temperature-dependent Fermi surface evolution [39]. Similarly, Lifshitz transitions and magnetic breakdown, processes that abruptly modify FS topology, are irrelevant here [38, 40]. The temperature-dependent evolution of $k_F$ in $EuCo_2Al_9$ aligns with exchange splitting modulated by magnetic order. As shown in Fig. 2**b**, the magnetization *M* grows with cooling, inducing band splitting ($\Delta_{ex}$) proportional to $I_{ex}\langle M_c\rangle$ [41–43]. This splitting arises from strong exchange coupling between localized Eu 4*f* and itinerant 5*d* electrons. Importantly, this coupling may also mediate the RKKY interactions between localized moments in the inversion-symmetric crystal lattice, stabilizing chiral spin textures [44].

## Discussions and conclusions.

Similar large Hall responses have been observed in other systems; for instance, in the Dirac semimetal $Cd_3As_2$, ultrahigh carrier mobility gives rise to a low-field peak in the Hall conductivity $\sigma_{xy}$ [45]. Based on first-principles calculations and quantum oscillation measurements, four Fermi pockets are identified as the origin of the observed quantum oscillatory transport. The relatively small amplitude of these oscillations compared to the total MR indicates that the Fermi surfaces contributing to the oscillations do not dominate the overall charge transport. This can be attributed to their large extremal cross-sections, such as the extensive, nearly cylindrical Fermi surface near the Γ point, which accommodates a substantial portion of the carriers. Consequently, we analyzed the high-field Hall conductivity using a two-band model. The fitting results reveal that the conduction is governed by two hole-like bands, with a total carrier concentration of approximately $2.1\times10^{22}$ cm$^{-3}$ ($n_{h1}$ = $1.15\times10^{22}$ cm$^{-3}$, $n_{h2}$ = $0.98\times10^{22}$ cm$^{-3}$) and nearly identical mobilities of about ~60 cm$^2$V$^{-1}$S$^{-1}$ for both pockets, at 2 K. The obtained mobility is substantially lower than the ultrahigh mobilities characteristic of $Cd_3As_2$ (~$10^6$ cm$^2$V$^{-1}$S$^{-1}$).

Recent studies have reported giant intrinsic AHE with large AHA in magnetic topological materials: GdPtBi exhibits a large AHA of 16–32 % [9, 46], while TbPtBi shows an even more pronounced AHA of 38–76 % [47, 48], which has been attributed to a magnetic-field-induced spin texture that drives an avoided band crossing. Despite these large AHAs, the corresponding AHC in both systems remains on the order of $10^2$ $\Omega^{-1}$cm$^{-1}$ [9, 46–48]. Although $EuCo_2Al_9$ also features a field-induced avoided band crossing, it delivers an extremely large AHC surpassing $10^4$ $\Omega^{-1}$cm$^{-1}$—orders of magnitude larger than those observed in GdPtBi and TbPtBi.

Recently, the emergence of a giant AHE driven by spin chirality skew scattering has been theoretically predicted and subsequently confirmed in a limited number of material systems [20–28]. In $EuCo_2Al_9$, based on magnetization and zero-field neutron diffraction measurements, we infer the possible emergence of a field-induced chiral spin structure on the triangular lattice of Eu moments. We note that the scalar spin chirality was not directly confirmed by performing neutron diffraction in our study. However, its presence is inferred and supported by the giant magnitude of AHC, the remarkably large AHA, combined with the frustrated magnetic lattice. This inference is further supported by quantum oscillations and first-principles calculations, which confirm the presence of RKKY interactions—a key condition for stabilizing such chiral spin configurations.

In conclusion, $EuCo_2Al_9$ constitutes a geometrically frustrated magnet with Eu forming a triangular lattice. It undergoes a magnetic transition at a temperature around $T^*$~3.5 K, resulting in an AFM state, while the magnetic structure below the lower transition at $T^\dagger$ requires further investigation. The temperature-dependent quantum oscillation analysis reveals substantial exchange splitting—a direct manifestation of strong hybridization between localized Eu 4*f* electrons and itinerant 5*d* electrons. This hybridization likely mediates an indirect RKKY interaction that stabilizes chiral spin textures through competing exchange pathways, leading to an extraordinary AHC of $3.1\times10^4$ $\Omega^{-1}$cm$^{-1}$ accompanied by an AHA approaching 12 %. This work establishes a new class of frustrated magnets as a platform for engineering the AHE, and elucidates the fundamental role of exchange interactions and spin textures in quantum transport. These insights lay a blueprint for unconventional spintronics that exploit emergent spin-texture dynamics.

## Method

### Crystal growth and magnetotransport measurements

Single crystals of $EuCo_2Al_9$ were synthesized via a flux growth method. A stoichiometric mixture of Eu:Co:Al (1:2:20 molar ratio) was sealed in an alumina crucible within a quartz ampoule under vacuum. The ampoule was heated to 1150 °C, homogenized for 10 hours to ensure complete melting and mixing, then slowly cooled at 3 °C/h to 750 °C to facilitate crystal growth. Excess molten Al flux was decanted via centrifugation at 750 °C, yielding hexagonal prismatic single crystals. Phase purity and composition were verified by energy-dispersive X-ray spectroscopy (EDX) and powder X-ray diffraction (XRD; PANalytical Empyrean diffractometer, Cu Kα radiation). Electrical transport and magnetization measurements were performed in Quantum Design Physical Property Measurement System (PPMS-14 T) with a vibrating sample magnetometer (VSM) option and Quantum Design Magnetic Property Measurement System (PPMS-5 T).

### Neutron diffraction

Neutron powder diffraction measurements were performed on the High Intensity Powder Diffractometer at China Advanced Research Reactor. Due to the large neutron absorption cross section of the Eu element, the $EuCo_2Al_9$ powders were mixed with Al powders in a mass ratio of 1:2. Then, the mixture was loaded into a vanadium can with a diameter of 3 mm. A neutron wavelength of 1.478 Å was used.

### Band structure calculations

The first-principles calculations of $EuCo_2Al_9$ were performed by using Vienna Ab-initio Simulation Package (VASP) [49, 50] based on the density functional theory (DFT) [51, 52]. The generalized gradient approximation (GGA) of Perdew-Burke-Ernzerhof (PBE) type [53] was chosen for the exchange-correlation functional. The kinetic energy cutoff of the plane-wave basis was set to be 350 eV. A 10×10×13 k-point mesh was used for the Brillouin zone (BZ) sampling. The Gaussian smearing method with a width of 0.05 eV was employed for the FS broadening. The spin-orbit coupling effect was included. In the calculations whose pseudopotential include the Eu 4*f* configuration, a GGA+U formalism of Dudarev et al. was used for correcting the correlation effect on Eu 4*f* electrons [54]. An effective Hubbard U was empirically set as 6 eV. The FSs were plotted by the Wannier orbital interpolation technique as implemented in Wannier90 package [55]. The intrinsic anomalous Hall conductivity was calculated with the Kubo-formula approach in the linear response scheme via the Berry curvature [2], as implemented in the WannierTools package [56], which is also interfaced to the Wannier90. A dense k-point mesh of 301×301×301 was employed using the Wannier interpolation method.

## Author contributions

Author contributions. Z.-A.X. coordinated the project. S.X. and J.-F.Z. contributed equally to this work. S.X., S.X.L., J.-J.M., T.-H.L., and Z.L. synthesized the single crystals with the assistance of J.M., W.-H.J., and X.F.X. J.-F.Z. and Y.X. performed the ab initio calculations. H.G., X.M., and W.Y. performed the neutron diffraction measurements. S.X. performed the magnetotransport measurements with the assistance of J.-F.L., Q.T., and Z.Z. S.X., H.G., and J.-F.Z. plotted the figures and analyzed the data. S.X., H.G., J.-F.Z., T.-L.X., and Z.-A.X. wrote the paper. All authors discussed the results and commented on the manuscript.

## Acknowledgments

Acknowledgments. This work was supported by the National Natural Science Foundation of China (Grant No. 12574177; 12174334; 12204410; 12074425), the National Key R&D Program of China (Grant No. 2024YFA1409002), the Innovation Program for Quantum Science and Technology (Grant No. 2021ZD0302500), the Natural Science Foundation of Zhejiang Province (LMS25A040002), the Fundamental Research Funds for the Central Universities, the Research Funds of Renmin University of China (No. 23XNKJ22), the Beijing National Laboratory for Condensed Matter Physics (2023BNLCMPKF008), the Guangdong Basic and Applied Basic Research Foundation (Grant No. 2022B1515120020), and the Chinese funding sources applied via HPSTAR.

## Data Availability

The data that support the findings of this study are available from the corresponding author upon reasonable request.

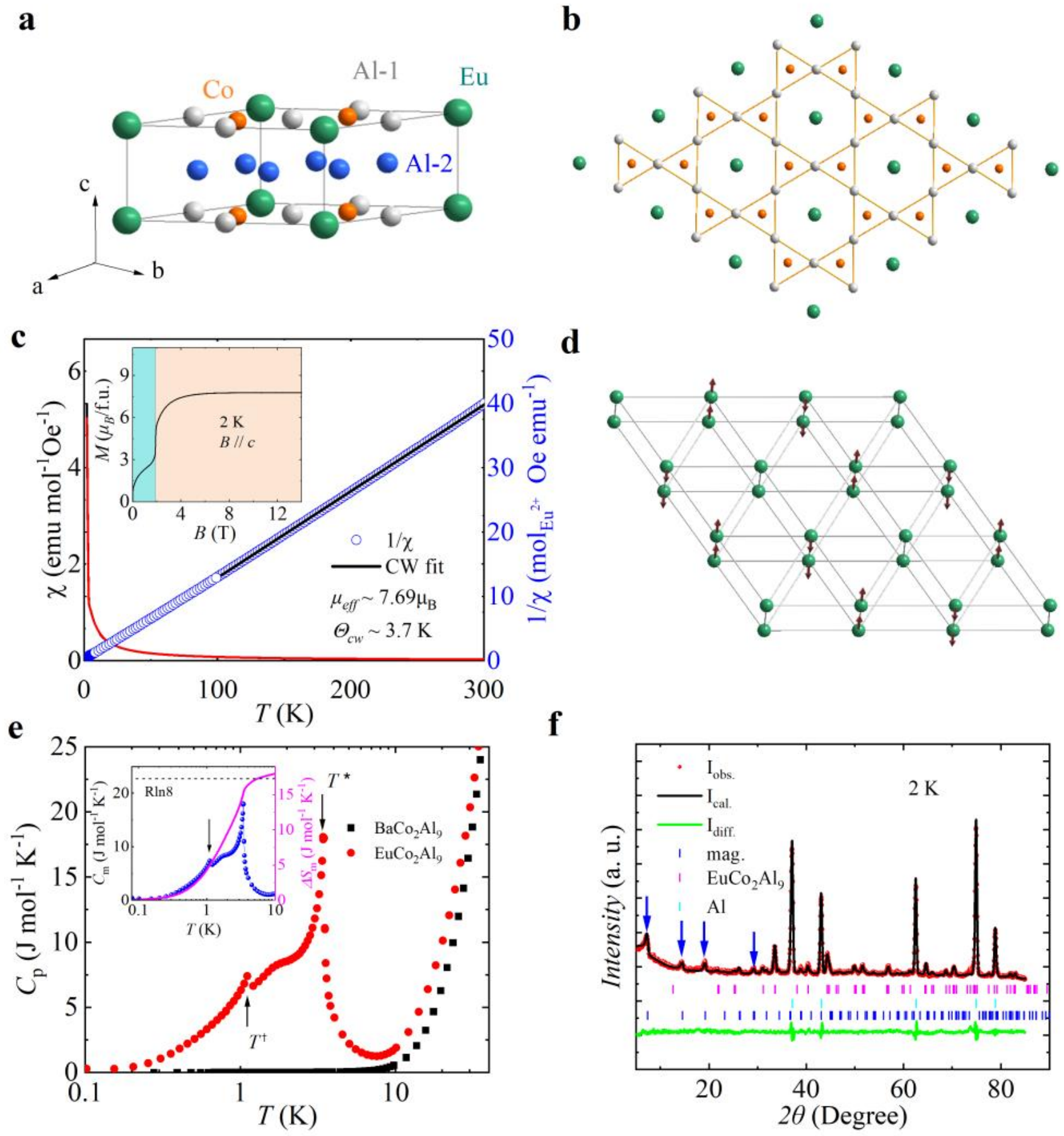

FIG. 1. **Crystal structure and magnetic phase transitions in $EuCo_2Al_9$. a,** Crystal structure. **b,** Atomic arrangement of the Eu-Co-(Al-1) sublattice in the *ab*-plane viewed along the *c*-axis. **c,** Temperature-dependent magnetic susceptibility (χ) and inverse susceptibility (1/χ). Solid black line: Curie-Weiss fit. Inset: Field-dependent magnetization at 2 K. **d,** Magnetic structure at 2 K and zero field. **e,** Heat capacity (main panel) of $EuCo_2Al_9$ and $BaCo_2Al_9$ at low temperature. Inset: Magnetic heat capacity and magnetic entropy of $EuCo_2Al_9$, separated by subtracting the phonon contribution based on the non-magnetic isostructural analogue $BaCo_2Al_9$. **f,** Neutron powder diffraction patterns at 2 K.

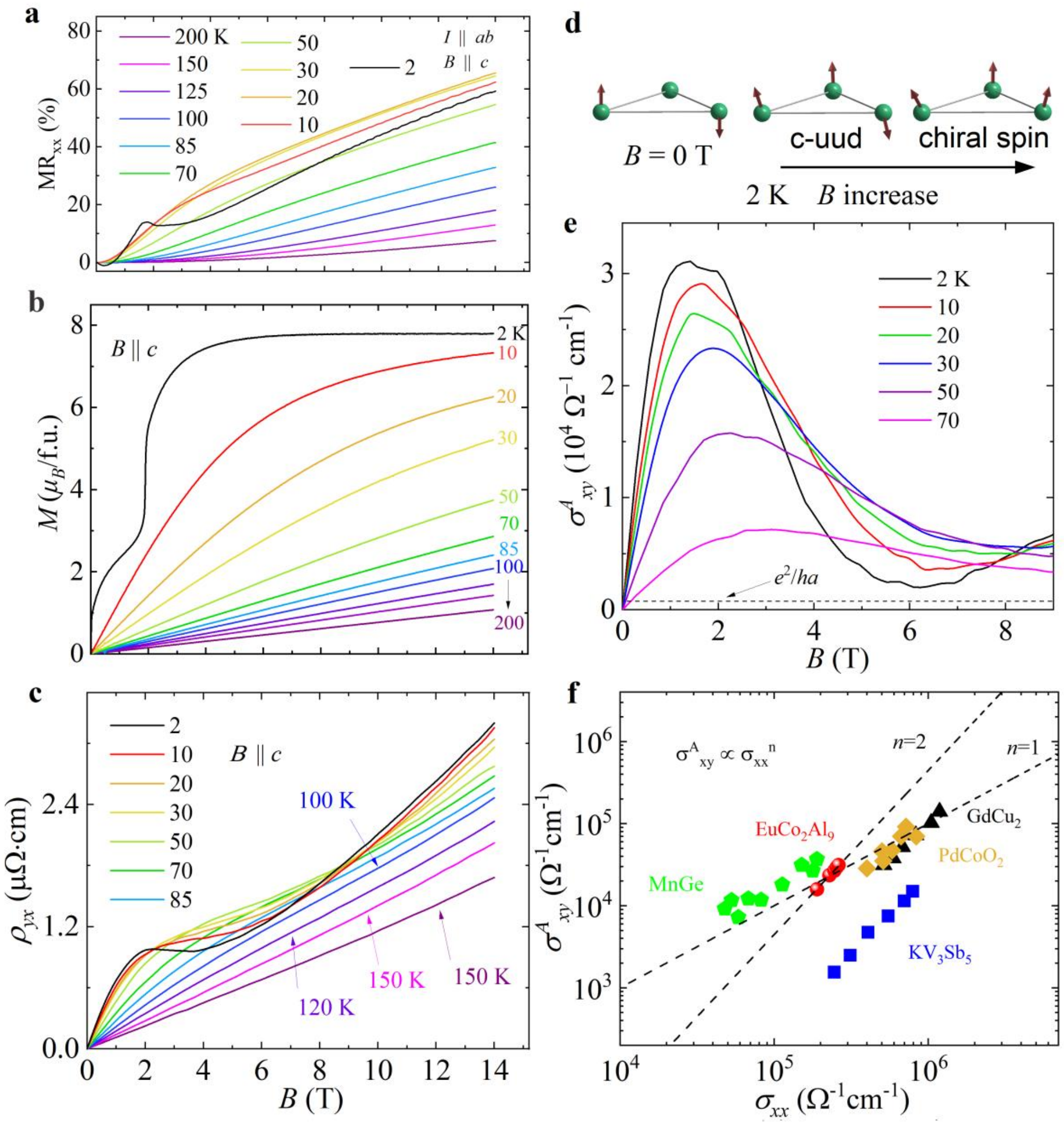

FIG. 2. **Observation of the giant AHE in $EuCo_2Al_9$.** **a,** Field-dependent magnetoresistance ($MR = (R_B - R_0)/R_0 \times 100\%$) at various temperatures. **b,** Field-dependent magnetization with $\boldsymbol{B} \parallel c$. **c,** Field-dependent Hall resistivity with the field applied along c-axis at different temperatures. **d,** Magnetic structure on the Eu triangular lattice at 2 K under different magnetic fields. **e,** Field-dependent AHC $\sigma^A_{xy} = \rho^A_{yx} / ((\rho^A_{yx})^2 + (\rho_{xx})^2)$ at different temperatures. f, Scaling relation $\sigma^A_{xy} \propto \sigma_{xx}{}^n$ for $EuCo_2Al_9$ and other spin chirality skew scattering materials $KV_3Sb_5$ [22], MnGe [23], $PdCoO_2$ [27], and $GdCu_2$ [28].

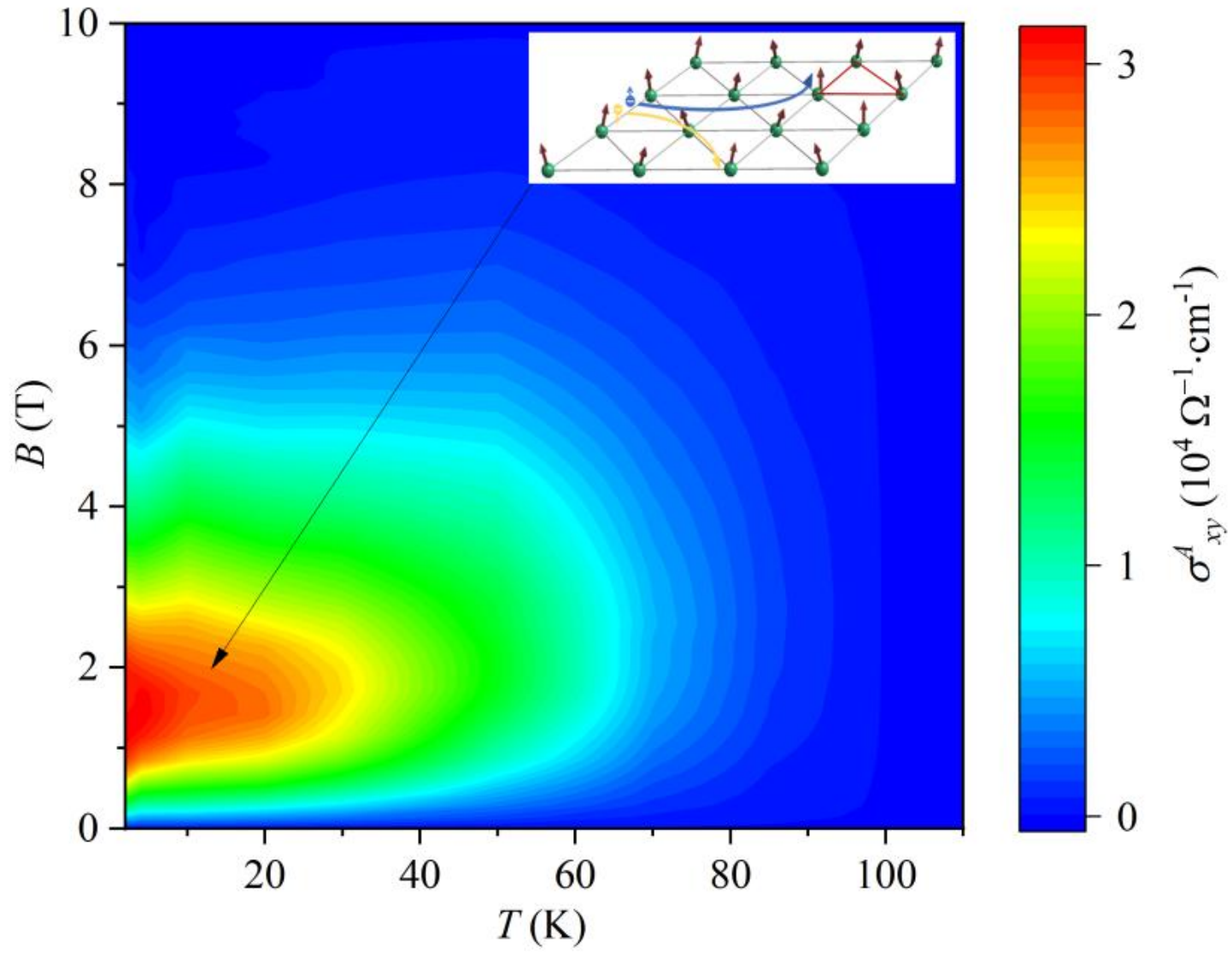

FIG. 3. **Contour plot of $\sigma^{A}_{xy}$ in the *B*−*T* space.**

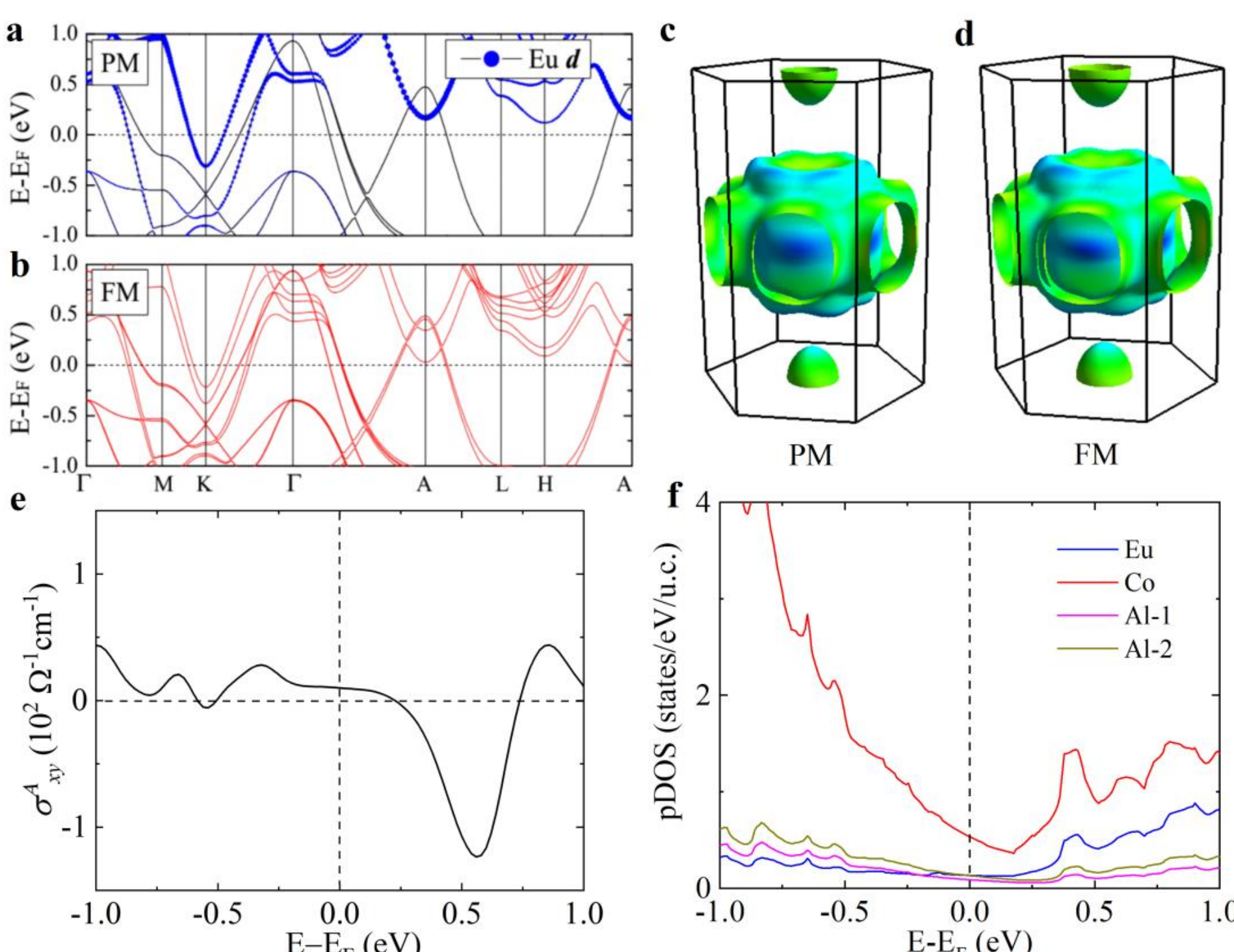

FIG. 4. **Electronic structure in $EuCo_2Al_9$. a,** Calculated band structures under the paramagnetic state. the sizes of blue dots are proportional to the weights of Eu atomic *d*-orbital. **b,** Calculated band structures under the ferromagnetic state. **c, d,** Fermi surfaces in the paramagnetic and ferromagnetic states, respectively. **e,** Calculated intrinsic AHC in the ferromagnetic state. **f,** Partial electronic density of states (pDOS) analysis around the Fermi level.

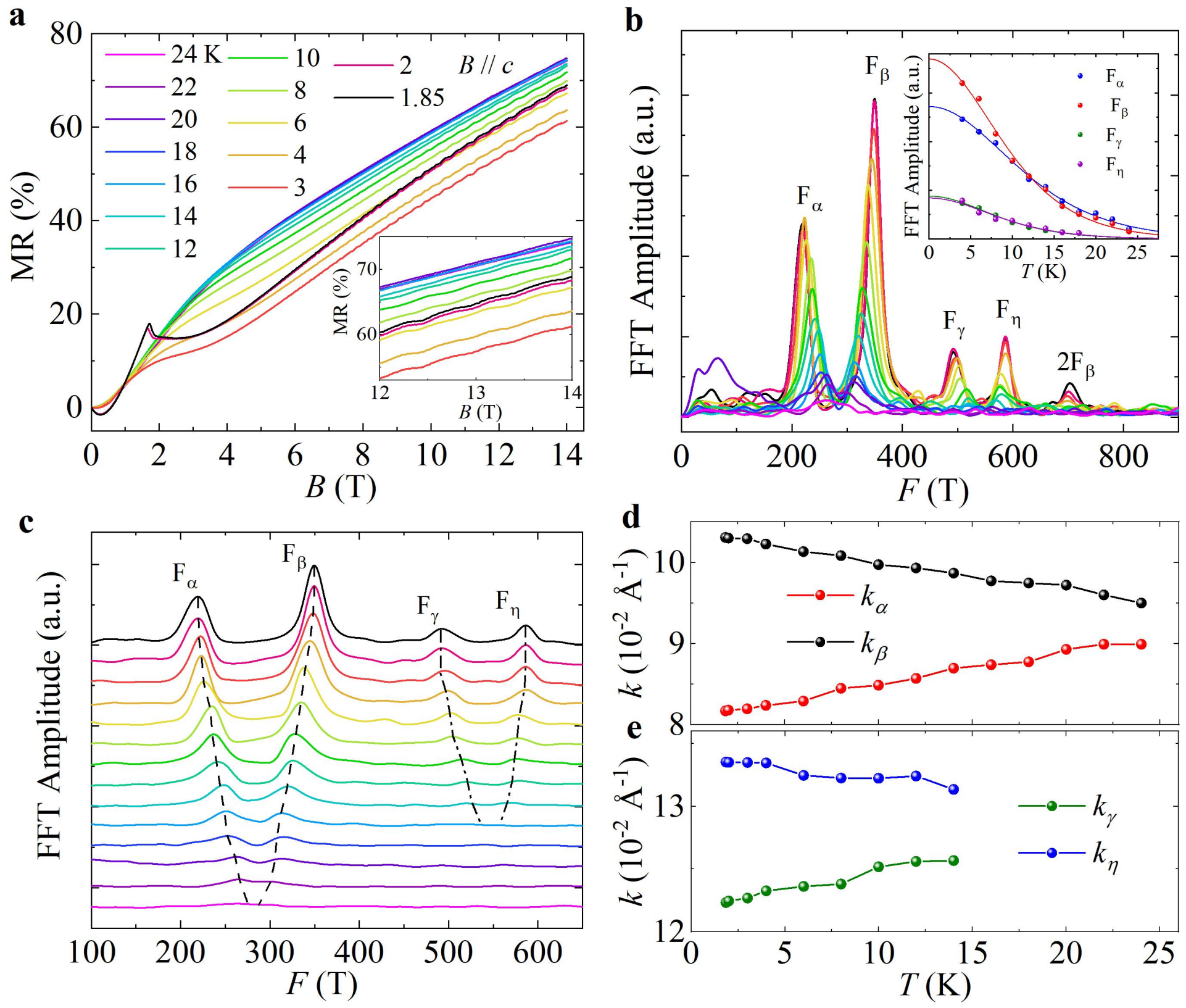

FIG. 5. **Shubnikov–de Haas (SdH) oscillations in $EuCo_2Al_9$.** **a,** SdH oscillations ***B*** // *c*. **b,** FFT spectra of $\Delta\rho_{xx}$. Inset: Temperature dependence of relative FFT amplitudes of the main oscillation frequencies. **c-e,** Temperature dependence of frequencies and the corresponding Fermi vectors, respectively.